# Magnitude of the Current in Two-Dimensional Interlayer Tunneling Devices

Randall M. Feenstra,[1] Sergio C. de la Barrera,[1] Jun Li,[1] Yifan Nie,[2] and Kyeongjae Cho[2]

[1]Dept. Physics, Carnegie Mellon University, Pittsburgh, Pennsylvania, U.S.A.

[2]Dept. Materials Science and Engineering, The University of Texas at Dallas, Dallas, Texas, U.S.A.

**Abstract:**

Using the Bardeen tunneling method with first-principles wave functions, computations are made of the tunneling current in graphene / hexagonal-boron-nitride / graphene (G/h-BN/G) vertical structures. Detailed comparison with prior experimental results is made, focusing on the magnitude of the achievable tunnel current. With inclusion of the effects of translational and rotational misalignment of the graphene and the h-BN, predicted currents are found to be about 15× larger than experimental values. A reduction in this discrepancy, to a factor of 2.5×, is achieved by utilizing a realistic size for the band gap of the h-BN, hence affecting the exponential decay constant for the tunneling.

## I. Introduction

Since the advent of microelectronic fabrication using layered two-dimensional (2D) materials more than a decade ago,[1] there have been a number of novel 2D devices produced. One of these in particular, the graphene/hexagonal-boron-nitride/graphene (G/h-BN/G) interlayer tunneling device,[2] has produced rather spectacular characteristics of tunnel current vs. bias voltage across the source and drain electrodes ( $I$ vs. $V_{DS}$).[3,4,5] Large negative differential resistance (NDR) is observed, with potential applications in oscillators and other types of circuits. This characteristic is an example of 2D-2D tunneling,[6] a phenomenon that has been studied for more than 25 years in a different 2D system, quantum wells in epitaxial layers of three-dimensional semiconducting systems such as GaAs/Al$_x$Ga$_{1-x}$As.[7,8,9,10]

A variant on the G/h-BN/G device has been proposed in which semiconducting 2D materials such as a transitional metal dichalcogenide (TMD) or phosphorene are used for the source and drain, rather than graphene.[11,12,13] A number of experimental efforts to produce such devices have been made, although the quality of the results considerably lag those in the G/h-BN/G system.[14,15,16] Nevertheless, it is important to consider the TMD-based devices, since they offer a wider range of operating modes with potential application in electronic circuits and systems.[17] Specifically, the TMD-based interlayer tunneling devices can be utilized for 2D-2D tunneling, just as for G/h-BN/G, and sharp resonant peaks exhibiting NDR are expected (though not yet experimentally observed).[18] In this mode, the tunneling occurs between *like bands*, that is, from conduction band (CB) of one electrode to CB of the other and/or from valence band (VB) of one electrode to VB of the other. Additionally, for the semiconducting TMDs, interesting characteristics are obtained when tunneling between *unlike bands*, i.e., VB of one electrode to CB of the other, or vice versa. This mode of unlike-band tunneling is identical to what is normally referred to as band-to-band tunneling in conventional tunneling field-effect transistors (TFETs),[19] bearing in mind that momentum conservation enters the problem in a two dimensional way for interlayer tunneling as compared to just a one dimensional manner for in-plane tunneling.

Due to the zero (or very small) band gap between VB and CB in graphene, the band-to-band type of tunneling in a G/h-BN/G device yields simply a linear $I$ vs. $V_{DS}$ characteristic.[6] In contrast, for a TMD-based device, band-to-band tunneling yields a sharp turn-on in the $I$ vs. $V_{DS}$ characteristic, when the VB edge of one electrode rises above CB edge of the other electrode.[11,12,17,18] (In typical TFET operation, the more relevant characteristic would be $I$ vs. gate voltage, but for the purpose of the present work it is sufficient to consider simply $I$ vs. $V_{DS}$.) This sharp turn-on permits usage of TFETs in circuits with reduced operating voltages, hence yielding digital devices and circuits that require relatively low amounts of energy to switch between *on* and *off* states.[19,20] In addition to this low-energy aspect of the devices, another critical requirement of TFETs is the amount of current that can be achieved in the *on* state. Typical values of 200 µA per µm length of channel are required for realistic circuit applications (for interlayer TFETs, this current is proportional to the width of the overlap region between source and drain, i.e. the channel width, which in our simulations is typically assumed to be 15 nm).[17]

A number of recent publications have presented simulation results for interlayer 2D TMD-based TFETs, utilizing either thin h-BN barriers or no h-BN at all (i.e. with only a van der Waals gap separating source and drain).[11,12,13,17,21,22,23] Relatively high current levels for the *on* state of the devices are obtained. However, as mentioned above, experimental confirmation of these tunnel current magnitudes is lacking. Nevertheless, since the theory for TMD-based and graphene-based devices is essentially the same, we can use the experimental results for the relatively well-defined G/h-BN/G devices[3,4] (electrodes that are single monolayer and rotationally aligned, together with precisely known barrier thickness) as a means of comparison for the theories, focusing on the magnitude of the tunnel current in particular. This type of comparison has been discussed in several recent works, examining in particular the possible effects of rotational misalignment between the h-BN layers and the graphene.[24,25] In our work we examine this same topic, using an approach based on first-principles wave functions and the Bardeen tunneling method[17] which we feel is somewhat more quantitative than that used in these prior works and hence permits a more detailed comparison with experiment.

For comparison with theory, we employ the recent experimental results of Mishchenko et al.[4] For a device with 4 layers of h-BN and known rotational misalignment between the two graphene electrodes (1.8°), a peak current of 0.27 µA/µm² at a source-drain voltage of 0.83 V can be deduced from Fig. 2(a) of Ref. [4] for bottom gate voltage of 0 V and no top gate. In order to permit direct comparison with our first-principles theoretical results, we make two adjustments to these experimental values: we modify them such that they correspond to a device with both zero graphene-graphene misorientation and zero graphene-graphene and graphene-gate capacitances (employing a specified, fixed doping density in the electrodes in place of the carrier density induced by capacitive effects).

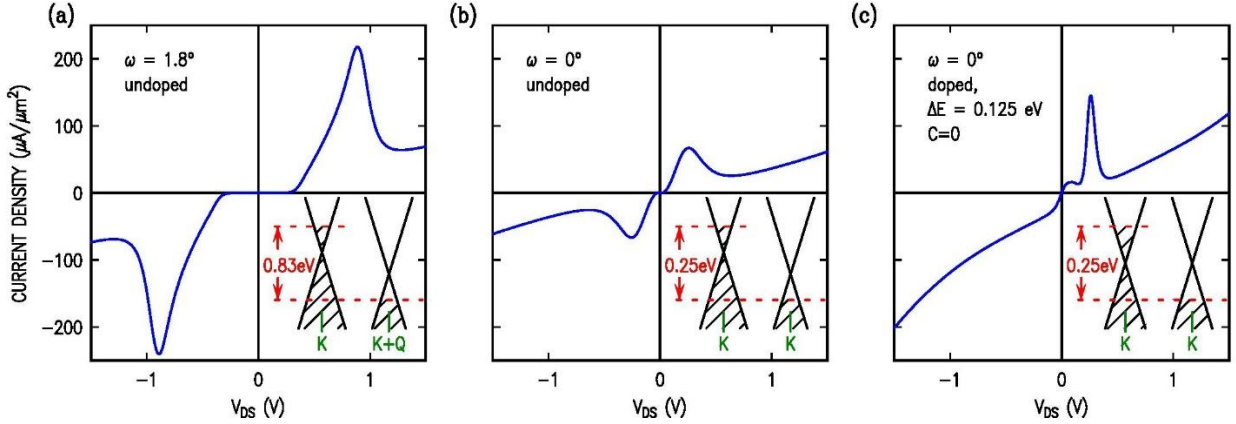

**FIG 1.** Simulation results using an analytic theory (Ref. [6]) for the G/h-BN/G tunneling device described in Ref. [4], for graphene-graphene misorientation and doping of (a) 1.8°, undoped; (b) 0°, undoped; (c) 0°, fixed doping corresponding to Fermi energies of ±0.125 eV (relative to Dirac point) in each graphene electrode. Band diagrams, corresponding to a source-drain voltage at the maximum of the resonant peak, are shown in each case. The bands are centered about the $K$ point in momentum space for (b) and (c) and also for the left-hand graphene electrode of (a), but for the right-band electrode of (a) the graphene bands are centered about a point $K + Q$ where $Q$ arises from rotational misorientation of the two electrodes.

Both the graphene-graphene misorientation and the electrostatic, capacitive effects in the device are fully described within a previously developed simulation method for G/h-BN/G devices,[6,26] as illustrated in Fig. 1. Hence we can use that theory to make the two adjustments just mentioned. We note that this prior method makes a number of assumptions regarding the vertical part of the wave functions extending out from the graphene electrodes, namely, that they are simply composed of decaying exponentials with assumed values for the decay constant and the prefactor of the exponential. These assumptions will certainly affect the magnitude of the tunnel current as obtained with this theory, but they are not expected to affect the predictions of misorientation or capacitive effects (since those effects rely on parts of the theory other than the vertical dependence of the wave functions). In any case, the prior theory permits evaluation of the current by a relatively straightforward numerical evaluation of a double integral (Eq. (10) of Ref. [6]); aside from that integral, all other expressions in the theory are evaluated analytically, and for lack of a better name we refer to the prior method as an "analytic theory".

Figures 1(a) and 1(b) show simulation results from the prior "analytic" theory for the Mishchenko et al. device with 1.8° and an equivalent device with 0° misorientation, respectively, including a full description of the electrostatics. The former results are very similar to those displayed by Mishchenko et al. (since the analytic theory that they use is essentially identical to ours).[4,6,26] For 1.8° misorientation the peak current is found to be 217 µA/µm² at a source-drain voltage of 0.83 V, whereas for 0° misorientation the peak current is 67 µA/µm² at a voltage of 0.25 V. This change of in the peak current resonant voltage due to the graphene-graphene misorientation is to be expected, since a finite twist angle between the graphene sheets has the effect of shifting the voltage by $\hbar v_F Q/e$ where the wavevector $Q$ extends between the $K$-valleys of the source and drain electrodes in momentum space[6] (also, an additional voltage shift occurs because of the larger graphene-graphene capacitance at the higher voltage). The influence of device electrostatics is illustrated in Fig. 1(c), where we have neglected all capacitances in the device and we have assumed a fixed carrier density (e.g. from doping) in each electrode corresponding to a separation between Fermi energies and Dirac point of 0.125 eV. Note that in this case a resonant peak is obtained for only one sign of the source-drain voltage, whereas in Figs. 1(a) and 1(b) peaks appear for both signs of the voltage. The difference between the situations can be explained with reference to the band diagrams shown as insets; in Fig. 1(c) the bands shift rigidly with the applied source-drain voltage, with the Dirac points being coincident at resonance, but for Figs. 1(a) and (b) the bands shift relatively slowly with applied voltage and the Dirac points are coincident at zero source-drain voltage (the energy difference between Dirac points in Figs. 1(a) and 1(b) is exaggerated in the diagrams, for clarity). The peak current in Fig. 1(c) is found to be 145 µA/µm$^2$ which is 0.67× smaller than in Fig. 1(a).

Returning to the experimental results of Mishchenko et al., we adjust their peak current of 0.27 µA/µm$^2$ such that it applies to a device with zero graphene-graphene misorientation and zero capacitance, thus yielding a current of 0.27×0.67 = 0.18 µA/µm$^2$ (at a source-drain voltage of 0.25 V). We employ this value for the adjusted, experimental current throughout the remainder of this paper, with the goal of our work being to compare that value with what we obtain from a first-principles (parameter free) theory. We note that the simulation results of Fig. 1(c) themselves provide a value for the peak current, 145 µA/µm$^2$, which is about 800× larger than the adjusted experimental value of 0.18 µA/µm$^2$. These simulation results of Fig. 1 are dependent on a number of parameters in the prior, analytic theory.[6,26] For one thing, certain wave function magnitudes have assumed values, i.e. parameters $u_{11}$ and $u_{12}$ of Ref. [6] are both set to unity. More importantly, a critical parameter is the exponential decay constant of the tunneling current as a function of the h-BN barrier thickness. We write this exponential decay as $\eta \exp(-2\kappa d)$, where $d$ is the thickness of h-BN, $\eta$ is a prefactor, and $\kappa$ is the tunneling decay constant (the factor of 2 in the exponent is included since, in the Bardeen tunneling approach, the current has contributions from wave functions extending out from *both* sides of the barrier, i.e. $\kappa$ gives the decay constant for each of those wave functions). In the analytic results of Fig. 1 we have used $\kappa = 6.0$ nm$^{-1}$ which, as discussed in Section III, is the same as that obtained in experimental work of Britnell et al. for the case of nonresonant tunneling.[2] Our first-principles theory of Section III provides predictions for the value $\kappa$; the results vary somewhat depending on the G/h-BN alignment, but a midpoint of this range falls close to 6.0 nm$^{-1}$. We thus consider this 6.0 nm$^{-1}$ value to be a reasonable estimate of the $\kappa$ value that describes resonant tunneling in the G/h-BN/G system.[(i)] However, in Section IV we argue that a significantly larger value of $\kappa$ is needed to truly describe such tunneling.

---

[(i)] Also concerning $\kappa$, in a prior work we employed a value of 5.2 nm$^{-1}$ based on a tight-binding analysis.[26] This value, which in retrospect is much too small, produced relatively large tunnel currents. Computed tunnel currents in Fig. 5 of that prior work deviate from experiment[3] by a factor of about 10$^4$, significantly more than the 800× factor discussed in the present work; most of the difference between these factors arises from the relatively small $\kappa$ value employed in the prior work.

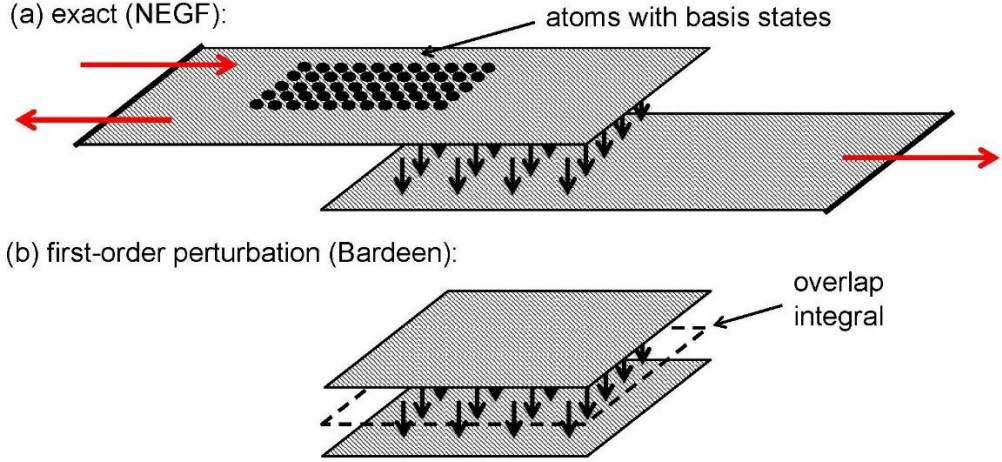

**FIG 2.** Comparison of geometries for computing vertical tunneling current between two 2D layers: (a) exact (or NEGF) computation, showing atom locations with approximate basis states on each atom, and (b) Bardeen method, using first-principles wave function of each electrode in the absence of the other.

Our theoretical method is described in Section II, with results provided in Section III. Section IV further discusses comparison between experiment and theory, and we also describe the relationship of our results with two recently published works dealing with the effects of misalignment between graphene and h-BN.[24,25]

## II. Method

We employ the Bardeen method for computing tunneling current, with all wave functions provided by first-principles density functional theory as implemented in the Vienna Ab-Initio Simulation Package (VASP),[27] with the projector-augmented wave method.[28] The Perdew-Burke-Ernzerhof form of the generalized-gradient approximation (GGA) for the density functional is used.[29] The wave functions are expanded in plane waves with a kinetic energy cutoff of 400 eV, and the convergence criterion for the electronic relaxation is $10^{-6}$ eV. Details of our computational method have been presented previously.[17] As an introduction to the Bardeen method,[30,31,32] we show in Fig. 2 a schematic illustration for comparing tunneling between sheets of 2D materials, as might be computed in an exact (or a non-equilibrium Green's function, NEGF) theory as compared with[33] a Bardeen method. In the former, as pictured in Fig. 2(a), a wave function (wave packet) approaches the tunnel barrier from the left-hand side. When the wave function reaches the area of the tunnel junction, then some of it will extend across to the opposite electrode, generating a transmitted wave that exits to the right-hand side of the lower electrode. Some of the incident wave is reflected, yielding the left-going wave in the upper electrode. In the NEGF method, this entire situation is modeled. However, such computations are not tractable using first-principles wave functions, so necessarily some approximate technique such as the tight-binding method must be employed in modeling the wave functions. This approximation then produces some uncertainty in the result.

With the Bardeen method, Fig. 2(b), the details of the incident and reflected waves do not enter into the theory. Rather, one employs the wave functions of electrode plus barrier, considered *separately* for the source and drain (upper and lower) electrodes. Transitions between these states are considered in a first-order perturbative sense, producing an expression for the current which contains an integral of the overlap of the wave functions over a plane midway between the electrodes. Hence, some details of the transport of the wave functions within the electrodes which *are* included in the NEGF method are *not* described within the Bardeen method. Nevertheless, the essential tunneling process is well described in the Bardeen method, and furthermore, since this method deals with the two separate subsystems of electrode plus barrier (and without the graphene leads that are adjacent to the barrier materials), then parameter-free first-principles computations are possible.

Our method follows precisely the one described in Ref. [17].[30,31,32,33] Briefly, the tunnel current is obtained from

$$I = \frac{4\pi e}{\hbar} \sum_{\alpha,\beta} |M_{\alpha\beta}|^2 [f_S(E_\alpha) - f_D(E_\beta)] \delta(E_\alpha - E_\beta) \tag{1}$$

with matrix element

$$M_{\alpha\beta} = \frac{\hbar^2}{2m} \int dS \left( \psi_\alpha^* \frac{d\psi_\beta}{dz} - \psi_\beta \frac{d\psi_\alpha^*}{dz} \right) \tag{2}$$

where $\alpha \equiv (\mathbf{k}_\alpha, \nu_\alpha)$ and $\beta \equiv (\mathbf{k}_\beta, \nu_\beta)$ label the states of the source and drain electrodes, having energies $E_\alpha$ and $E_\beta$, respectively, $m$ is the free electron mass, $f_S$ and $f_D$ are Fermi occupation factors for the electrodes, $f_S(E) = \{1 + \exp[(E - \mu_S)/k_B T]\}^{-1}$ and $f_D(E) = \{1 + \exp[(E - \mu_D)/k_B T]\}^{-1}$, where $\mu_S$ and $\mu_D$ are the chemical potentials in the two electrodes, $\mu_S - \mu_D = -eV_{DS}$, where $V_{DS}$ is the applied bias between source and drain. The surface integral in Eq. (2) is evaluated over the plane separating the electrodes. We have argued previously that a useful model for evaluating the surface integral is to restrict it to a phase coherent area for the wave functions in the respective electrodes, with extent given by a phase coherence length $L$.[6,17,26] Our method actually employs a distribution of such lengths, centered on the value $L$, following the methodology of Britnell et al.[3]

For the treatment of the electrostatics of the tunneling device we employ a method that is more approximate than that of Ref. [17] but nonetheless adequate for the present purposes. That is, we neglect any role of gate electrodes on either side of the source and drain and we furthermore neglect the role of the capacitance between the graphene source and drain electrodes. Hence, specification of the doping of the graphene source and drain electrodes (equal and opposite doping in opposing electrodes) then determines the density of carriers in each electrode according to $n = (\Delta E/\hbar v_F)^2/\pi$ where $v_F$ is the Fermi velocity for graphene and $\Delta E \equiv |\mu - E_D|$ is the specified separation between Fermi energy and Dirac point in each electrode (which, neglecting capacitance, is a fixed quantity, independent of $V_{DS}$). At resonance, the bias voltage between the electrodes is given by $|V_{DS}| = 2\Delta E/e$. All of our computations are for an assumed temperature of 0 K (these results for 2D-2D tunneling have relatively little temperature dependence, since, even with thermal occupation of carriers, wavevector conservation in the tunneling is dominant in determining the width of the resonant peak, see e.g. Fig. 6 of Ref. [6]).

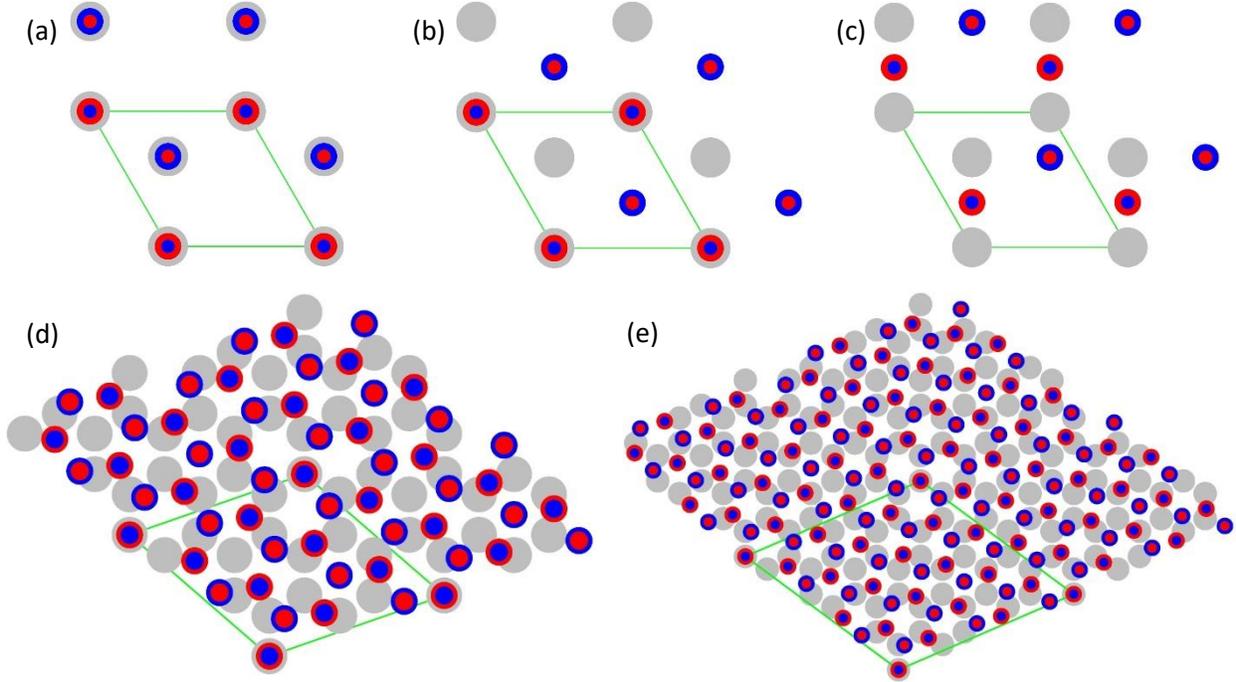

**FIG 3.** Atomic arrangements used for computations, showing one monolayer of graphene (gray filled circles) and two monolayers of h-BN (red and blue filled circles for B and N, respectively). Smaller circles correspond to atoms that are further above the graphene plane. Unit cells are shown in green. (a) – (c) 1×1 arrangements with 0° rotation between graphene and h-BN lattices and various translational registrations (aligned, Bernal, and bridging, respectively), (d) $\sqrt{7} \times \sqrt{7}$ with 21.79° rotation, and (e) $\sqrt{19} \times \sqrt{19}$ with 13.17° rotation.

We study various geometries of the graphene and h-BN in each electrode, as pictured in Fig. 3. For an assumed 1×1 unit cell between them, i.e. equal lattice constants and zero rotational misalignment, we study three different translational stacking arrangements as shown in Fig. 3(a) – (c), which we refer to as aligned, Bernal, and bridging, respectively. For a given G/h-BN stacking in one electrode, we are free to vary the stacking in the other electrode, so long as the h-BN material in the barrier assumes its usual AA' stacking sequence. We then obtain results for various different translational alignments for both the graphene and h-BN as well as between the two graphene electrodes.

Considering rotational misalignment of the h-BN and the graphene, we have considered two of the nearly lattice matched structures used by prior workers.[24,25,34] These structures are pictured in Fig. 3(d) and 3(e), having rotation angles between the graphene and h-BN of 21.79° and 13.17°, respectively, and with commensurate unit cell sizes of $\sqrt{7} \times \sqrt{7}$ and $\sqrt{19} \times \sqrt{19}$. Again, as for the 1×1 unit cells of Figs. 3(a) – (c), we are free to assume translational shifts of the graphene source and drain electrodes relative to each other (so long as the h-BN in the barrier maintains AA' stacking). For the $\sqrt{7} \times \sqrt{7}$ unit cell there are therefore 7 structures with different graphene-graphene translational alignment that are considered, and for the $\sqrt{19} \times \sqrt{19}$ unit cell there are 19 such structures.

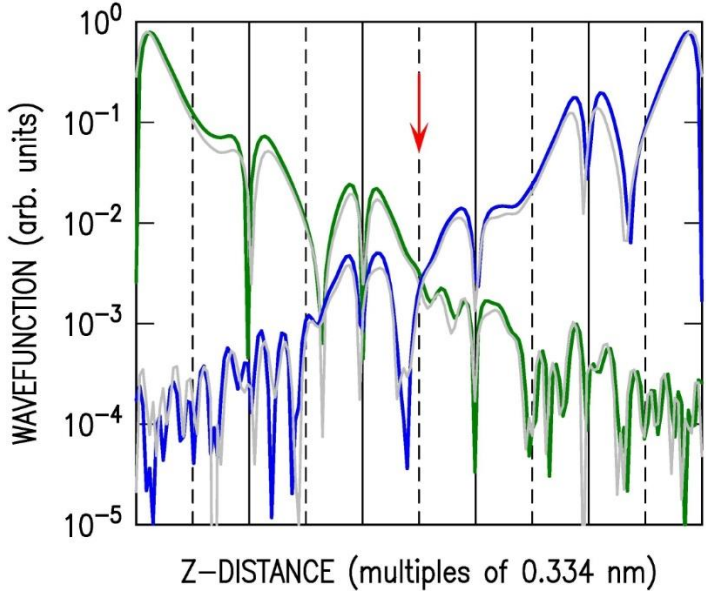

**FIG 4.** Wave functions at the Dirac point for 1×1-aligned case, extending out from the source and drain electrodes. Solid vertical lines show the locations of graphene or h-BN planes, and dashed vertical lines show the midpoints of interlayer spaces. For the case of four h-BN layers in the barrier, each of the wave functions is evaluated at a location between the second and third h-BN layers, as indicated by the red arrow. Blue and green curves show results for no relaxation of the atoms, whereas gray curves include relaxation.

In order to evaluate the tunnel current (Eq. 1), wave functions are evaluated along planes located midway between graphene and h-BN layers or midway between neighboring h-BN layers, in each electrode. All of our DFT results are obtained using a slab containing one graphene monolayer and four h-BN monolayers, surrounded by vacuum in a supercell with 3 nm periodicity perpendicular to the slab. Wave function amplitudes are shown in Fig. 4, for the 1×1-aligned case and evaluated at a lateral (x,y) location containing a C atom in the graphene plane and a B (N) atom in the first BN plane beside the left (right) electrode. For illustrative purposes we display the wave functions extending out from both electrodes into the h-BN; the matrix element term (Eq. 2) can be thought of approximately as the overlap of these wave functions, integrated over the entire plane separating the electrodes (and applying the current operator to the wave functions, as in Eq. 2). The choice of the specific plane is determined by the h-BN barrier thickness being considered; e.g. as pictured in Fig. 4 for the case of 4 layers of h-BN, the G/h-BN computations from each electrode are evaluated at a location between the second and third h-BN layers. As already mentioned above, our evaluations of the two respective G/h-BN computations are made such that the h-BN in the barrier properly has AA' stacking. Hence, for even numbers of h-BN layers, the respective computations for the two electrodes (for the same G/h-BN registration in the two electrodes) have the locations of the B and N atoms, relative to C, reversed.

For computational convenience, we do *not* relax the atomic positions in our computations, but rather, we maintain ideal positions within each plane (as for perfect monolayers of each material) with 1×1 unit cell size of 0.247 nm and with fixed interplanar separation of 0.334 nm. However, for the 1×1 aligned case we have also performed computations in which the atomic positions *are* relaxed. Wave functions including relaxation are shown in Fig. 4; it is apparent that they are only slightly affected by the relaxation, with the resulting tunnel currents being reduced by a factor of 2. The influence of the relaxation may be different for the rotationally and/or translationally misaligned situations (see discussion at end of Section V regarding moiré patterns), but we do not attempt to model the atomic relaxation of those situations. We also note in Fig. 4 that the appearance of the wave functions, with or without relaxation, become slightly irregular (noisy) at distances greater than about 4×0.334 nm from the graphene. Possibly this irregularity extends to the interlayer space at 3.5×0.334 nm, and hence the evaluation of tunneling currents that require wave function values on this place (i.e. for 5 or 6 h-BN layers in the barrier) might be somewhat limited in their precision. For our $\sqrt{19} \times \sqrt{19}$ computation we have also run it with a tighter convergence criterion of $10^{-7}$ eV, but we find that the tunnel currents vary by only a few % (increased precision for 5 or 6 h-BN layers would likely require a higher kinetic energy cutoff).

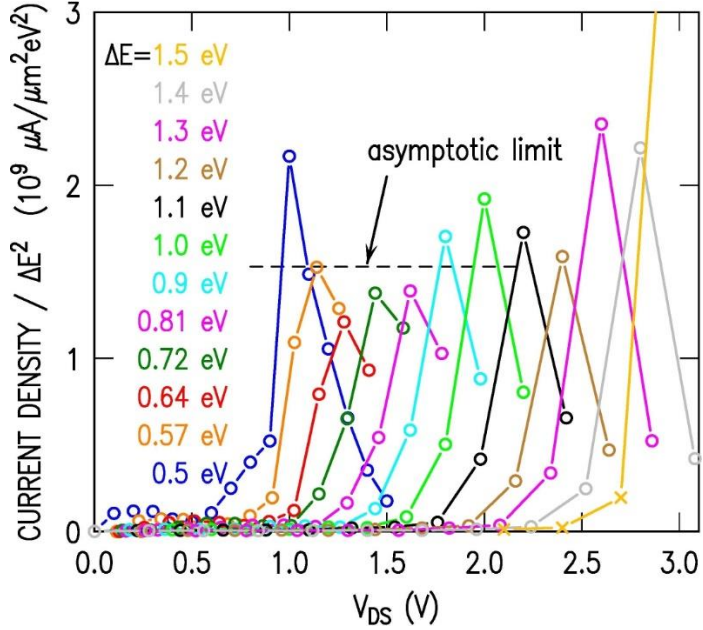

**FIG 5.** Computed tunnel current vs. source-drain bias voltage for various assumed doping concentrations (specified by $\Delta E$) in the source and drain, for a 1×1-aligned arrangement in both electrodes. An estimated asymptotic low-voltage limit (in the absence of grid-size effects) is indicated.

For the case of zero h-BN layers, the evaluation of the tunnel current as described above is somewhat ambiguous. We could use the wave function extending out from the graphene evaluated at the midpoints between the graphene and the nearest h-BN plane of our computed slab, although if we choose to do so then it's not clear whether or not we should utilize computations for the two electrodes with identical B (and N) locations in the nearest h-BN plane, or with reversed locations. Alternatively, we could use wave functions extending out from the graphene electrodes into vacuum (computed using isolated graphene surrounded by vacuum, i.e. with no h-BN). For the 1×1 aligned geometry, the maximum difference in tunnel current between these three situations is found to be only 11%, and to be definite we use the wave functions that extend into the h-BN and with reversed locations of the B and N locations in the nearest h-BN plane (hence, this case of zero h-BN layers follows precisely the same methodology as the other cases with even numbers of h-BN layers).

For our computations of the tunnel current, we employ Monkhorst-Pack **k**-space grids of size 36×36×1, 15×15×1, and 9×9×1 for the 1×1, $\sqrt{7} \times \sqrt{7}$, and $\sqrt{19} \times \sqrt{19}$ unit cells, respectively.

These meshes are somewhat limited in size by computational constraints, and hence even for relatively large assumed doping densities of the graphene sheets, the vast majority of the states that contribute to the tunnel current are located within a single grid point from the Dirac points of the respective band structures. We note that our method includes interpolation between neighboring **k**-points,[17] but nevertheless, with our somewhat coarse grids we still encounter some nontrivial grid-size effects. Band structures for 4 layers of h-BN on graphene for the 1×1, $\sqrt{7} \times \sqrt{7}$, and $\sqrt{19} \times \sqrt{19}$ arrangements are pictured in Fig. S1 of the Supporting Information. The energies of the states nearest to, but not at, the Dirac point are seen in those plots to be 0.65, 0.65, and 0.67 eV, respectively, although additional points (not along the *MK* direction in **k**-space) occur in each case at energies of 0.42, 0.38, and 0.38 eV, respectively. Assumed doping densities for our computations must yield Fermi energies (relative to Dirac point energy) that are significantly greater than these minimal energy separations in order that avoid serious grid-size effects. Figure 5 displays tunnel current computations for 1×1-aligned arrangement with zero h-BN layers, for various assumed Fermi energies as listed and for a lateral coherence length of 10 nm. All the curves seen there display a resonant peak located at a bias voltage of twice the Fermi energy, as expected.[6] We are plotting the tunnel current divided by $\Delta E^2$, i.e. dividing out the expected dependence of the magnitude of the resonant peak with doping density. Some oscillation in the resultant magnitude of the resonant peaks in Fig. 5 is seen; a local minimum for Fermi energy (relative to Dirac point) of 0.64 eV and a local maximum for 1.0 eV. Additionally, an overall increase in the magnitude of the peak occurs for increasing doping, since the respective energies of the relevant states also increase and hence reduced barrier heights occur for the wave functions; this effect is especially apparent for $\Delta E$ values above 1.0 eV ($V_{DS}$ above 2.0 V). The oscillations in the peak magnitude arise from finite grid-size effects. To approximately account for these effects, we estimate an asymptotic peak current for low Fermi energies (low $V_{DS}$), as indicated in Fig. 5, located midway between the local peak minimum and maximum magnitudes just mentioned. We perform all of our computations reported henceforth in this work with assumed Fermi energies relative to Dirac point of ±0.5 eV, but we then correct all of these results to account for the grid-size effects. This correction amounts to a factor of 0.71 for the 1×1 case (ratio of asymptotic limit in Fig. 5 and peak height for 0.5 eV Fermi energy), and with similar analyses we find correction factors of nearly the same value for the $\sqrt{7} \times \sqrt{7}$ and $\sqrt{19} \times \sqrt{19}$ cases. Hence, with our chosen **k**-space grid spacings, we achieve comparable accuracy in our computations for all of the various unit cells. Additionally, for the 1×1 unit cell we have employed a grid that is 16 times denser (restricting the grid to include only points near the K-point), as described in Fig. S2. As shown there, the results obtained from the coarse and fine grids are in reasonably good agreement.

In order to compare our theoretical results with the experimental results of Mishchenko et al.,[4] two additional corrections must be made. The theoretical computations as just described are performed with Fermi energy of 0.5 eV and coherence length of 10 nm. In contrast the adjusted experimental result (see Section I) has Fermi energy of 0.125 eV and coherence length of 50 nm. Correction factors to enable a one-to-one comparison of experiment and theory are thus $(0.125/0.5)^2$ for the doping, multiplied by (50/10) for the coherence length, yielding a factor of 0.3125. In our summary plots below of tunnel current that compare theory with experiment, we apply this correction to the theory.

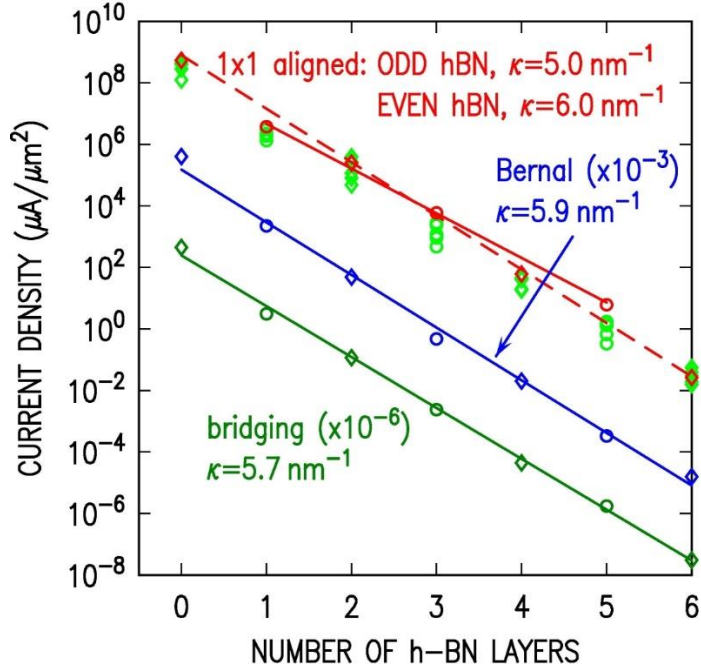

**FIG 6.** Tunnel current vs. number of h-BN layers, for 1×1 rotational alignment of graphene and h-BN and for various translational registrations between the h-BN and graphene. Aligned, Bernal, and bridging registry are shown by red, blue and dark green, respectively. Lines show fits to $\eta \exp(-2\kappa d)$, where $d$ is the thickness of h-BN, $\eta$ is a prefactor, and $\kappa$ is the tunneling decay constant. Light green symbols show all of these results together, and also include results for different G/h-BN translational registry in the two graphene electrodes.

### III. Results

Figure 6 shows the first-principles results for the tunnel current at 1 V source-drain bias (i.e. at the maximum of the resonant peak) for the 1×1 case with various translational registration of the graphene and h-BN stacking as indicated, as a function of the number of h-BN layers in the barrier. Examining the results, we see for the aligned stacking that different values for the decay constant $\kappa$ are obtained for even or odd numbers of h-BN layers in the barrier. However, this difference is not found for either the bridging or the Bernal stacking. We interpret this behavior in terms of the two different sublattices (pseudo-spin values) for the C atoms in the graphene. For the aligned stacking, the two sublattices are clearly differentiated based on the proximity of their respective C atoms to the B or N atoms of the h-BN, as in Fig. 3(a). For odd numbers of h-BN, then the *same* sublattices will be spatially aligned with each other between the two electrodes, whereas for an even number of h-BN layers it is *opposite* sublattices of the respective electrodes that are spatially aligned. Hence, the lateral parts of the wave functions will be partially orthogonal for the even-number-of-BN case, and hence the tunnel current decays faster than for the odd-number-of-BN case. In any case, it is important to note that in G/h-BN/G experiments we do *not* expect these sublattice effects to be relevant, since the 1.6% lattice mismatch between h-BN and graphene (along with the inability to exfoliate and transfer layers with atomic precision in placement) ensures that a more or less random atomic translational registration of the graphene and h-BN lattices will occur.

For the bridging case shown in Fig. 6, there is no difference between the two sublattices within the graphene, and hence only one decay constant is obtained. For the case of Bernal stacking of the graphene and h-BN, the situation is slightly more complicated. Again, the proximity to a B or N atoms will select one particular sublattice, so for odd numbers of h-BN atoms the situation is the same as for the aligned case. However, for even numbers of h-BN layers, we end up with a B or

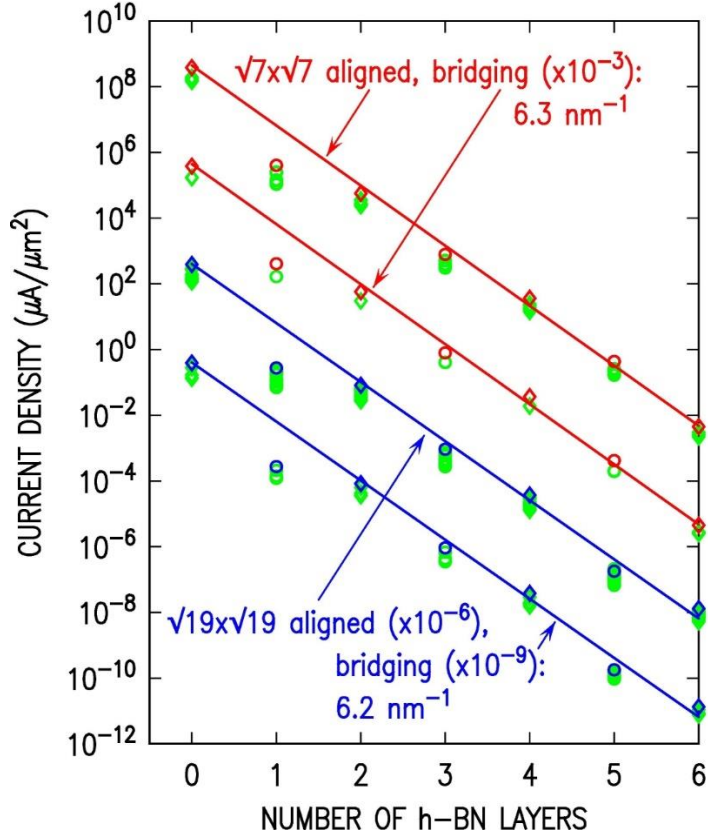

**FIG 7.** Tunnel current vs. number of h-BN layers, for $\sqrt{7} \times \sqrt{7}$ and $\sqrt{19} \times \sqrt{19}$ rotational alignments between the h-BN and graphene, and for aligned and bridging translational alignments in each case. Results for different G/h-BN translational alignments in the two electrodes are shown in light green. Lines shows fits to $\eta \exp(-2\kappa d)$, as in Fig. 6, but with the data points for barrier thickness of 1 h-BN layer being neglected in the fits.

N directly opposite C atoms that are spatially aligned (and with neither B nor N being near the C atom of the other sublattice). This arrangement ends up producing essentially no change in the decay constant as compared to the odd-number-of-BN case, since the *same* sublattices in each electrode end up being selected by the neighboring B or N atoms (i.e. even though the B and N are different atoms, their effect on the neighboring C atoms is relatively similar, as compared to what is experienced by the C atoms which do not have nearest neighbor B or N atoms).

Results for the $\sqrt{7} \times \sqrt{7}$ and $\sqrt{19} \times \sqrt{19}$ rotational alignments of the h-BN relative to the graphene are pictured in Fig. 7. The aligned translational registration shown there has one B or N atom in line with a C atom in each electrode, whereas the bridging registration has this B or N atom located precisely between two neighboring C atoms. For the results of Fig. 7, in contrast to Fig. 6, we now find nearly no difference between the aligned and bridging translational registrations. This result demonstrates that any sublattice symmetry (i.e. for the aligned cases) of these $\sqrt{7} \times \sqrt{7}$ and $\sqrt{19} \times \sqrt{19}$ situations is too small to produce a significant impact on the tunnel currents. Hence, the tunnel current vs. number of h-BN layers can be fit with a single $\kappa$ value for both even and odd number of h-BN layers, but only if we choose to neglect the current for a barrier thickness of 1 h-BN layer. Those currents are much reduced compared to the fit lines shown in Fig. 7. That is, the dependence of tunnel current on barrier thickness is found to be non-exponential, with a sharp drop in the current for 1 h-BN layer but with the current then recovering to values close to the fit lines for barriers consisting of 2 or more h-BN layers. Inspection of the wave functions for these $\sqrt{7} \times \sqrt{7}$ and $\sqrt{19} \times \sqrt{19}$ situations reveals a similar decrease in the wave function amplitude as a function of z-distance as for the 1×1 wave function of Fig. 3, i.e., without any additional drop in amplitude at the *z* locations of 0.5, 1.0, or 1.5 times 0.334 nm. Hence, the distinct drop in the current for 1 h-BN layer seen in Fig. 7 arises from a poor overlap between the *lateral* parts of the wave functions when computing the overlap at 0.5×0.334 nm for one electrode and 1.5×0.334 nm for the other. This poor overlap is not surprising, since the former wave function has its maxima over the locations of the C atoms and the latter over the locations of the B or N atoms, and there is poor alignment between the locations of these respective atoms. However, evaluation of currents for thicker barriers is based on evaluation of the wave function at locations exclusively between h-BN layers, so the overlap of the lateral parts of the wave functions recovers to similar (or slightly reduced) values as for the 1×1 situation.

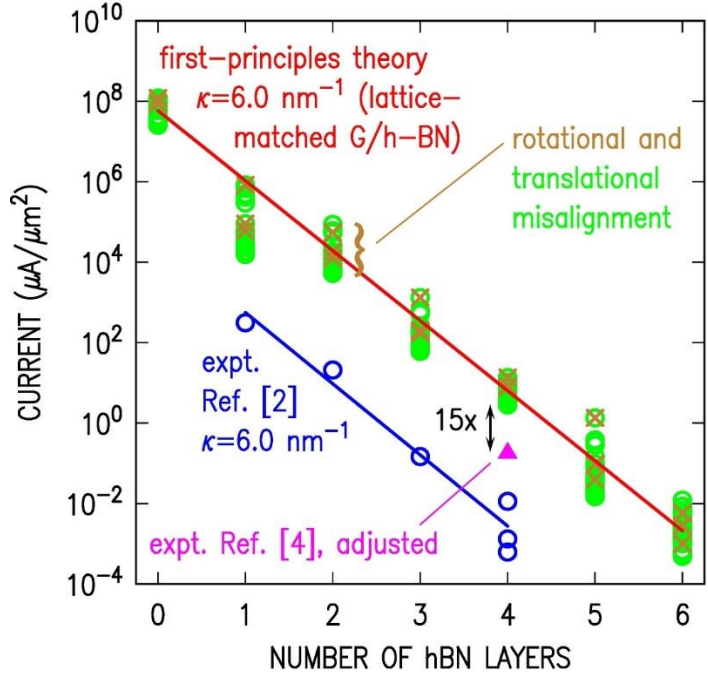

**FIG 8.** Summary of theoretical results, compared to experimental results from references indicated. For the theory, brown crosses show results for translationally aligned graphene electrodes and varying graphene/h-BN rotational alignment, whereas green circles include translational misalignment of the graphene electrodes. Lines show fits to $\eta \exp(-2\kappa d)$, as in Fig. 7. Comparing the theory (including misalignment) to the experiment of Ref. [4] for resonant tunneling, a discrepancy of 15× occurs between them.

To ascertain the overall decrease in tunnel current due to rotational and/or translational misalignment of the h-BN and the graphene, we show in Fig. 8 all of our first-principles results plotted together, with brown crosses used to show results for translationally aligned graphene electrodes and green circles for misaligned electrodes. For these results in Fig. 8 we have included the correction factor of 0.3125, as discussed in Section II, in order to make the theory results apply to a similar device configuration (doping, and coherence length) as the experimental result shown there. Overall, we see that the inclusion of the rotational and translational misalignments produces a significant drop in the computed tunnel current. This drop is most significant for 1 h-BN layer and it depends on the precise misalignment. As seen in Fig. 8, the currents for 1 h-BN layer vary by a factor of about 100, whereas for 4 layers they vary over only a factor of 5. For larger numbers of h-BN layers this variation appears to increase, although numerically we have some additional uncertainty in the results for 5 and 6 layers (since the wave functions, as pictured in Fig. 4, start to display some noise at the location between the third and fourth h-BN layers) so we do need not place too much significance on this latter result.

Overall, our theoretical results are consistent with a decay constant $\kappa$ of about 6.0 nm$^{-1}$, as indicated in Fig. 8. The experimental data of Britnell et al.[2] for nonresonant tunneling is also plotted in Fig. 8, and it also displays a similar decay constant of 6.0 nm$^{-1}$. For the results of Fig. 6 that strongly preserve sublattice symmetry a smaller value of $\kappa$ is found, 5.0 nm$^{-1}$, but this situation only occurs for very specific registration of the graphene and the h-BN, something that is exceedingly unlikely to occur in experiment. We therefore consider the computational values obtained in cases where sublattice symmetry is not preserved to be the most relevant ones, i.e. in the range 5.7 – 6.3 nm$^{-1}$ from Figs. 6 and 7, or with average value of about 6.0 nm$^{-1}$.

Let us examine the results in Fig. 8 for 4 h-BN layers, the barrier thickness used in experiment.[4] Considering the variation in computed tunnel currents that occurs due to translational misalignment of the two graphene electrodes (green circles), values on the high end of this range occur for perfectly aligned electrodes. Such electrodes will certainly not occur in the experiments (i.e. which employ exfoliated and transferred graphene flakes),[4] so we focus instead on the low end of the range of computed values. Compared to that, we find a factor of 15× discrepancy between theory and experiment, as indicated in Fig. 8. In the following section we explore the origin of this discrepancy.

## IV. Discussion

In this section we propose a reason for our remaining discrepancy of 15× between theory and experiment as shown in Fig. 8, and we also discuss the connection between the results achieved in the present work compared to two recent works that addressed the same topic,[24,25] However, before dealing with those topics, we return to consider the 800× discrepancy that was found in Section I when comparing the prior, analytic theory (with its assumed parameter values) to experiment. This discrepancy is a factor of $800/15 \approx 50×$ larger than for the first-principles results. The value of the decay constant $\kappa$ used in the estimates based on the analytic theory was 6.0 nm$^{-1}$, i.e. the same as deduced from the first-principles results of Fig. 8. Hence, it is not this parameter in the analytic theory that produces the relatively large discrepancy with experiment; rather, the source of the larger discrepancy must be the prefactor for the tunneling current. The analytic theory contains parameters $u_{11}$ and $u_{12}$ in the prefactor (in separate terms, which are summed together), both of which were assumed to have values of unity.[6] The final tunnel current depends on the fourth power of these factors, so apparently a better estimate for these factors would have been about $50^{-1/4} \approx 0.38$ (we make no effort to separately evaluate $u_{11}$ and $u_{12}$ here, with this 0.38 value applying to the maximum of the two). We note that a normalization parameter $D$ also appears in the prefactor of the analytic theory, with assumed value of the interlayer separation in graphite (0.335 nm). Modification of this value could also be made, with corresponding adjustment of the $u_{11}$ and $u_{12}$ values.

We can be somewhat more definite regarding the shortcomings of the analytic theory. To consider a specific situation, let us return the 1×1 aligned case described in Figs. 4 and 5. Although the wave functions of Fig. 4 display rather complicated behavior, their values at the locations of the interplanar midpoints (dashed lines in Fig. 4) do, in fact, follow an exponential decay fairly well (at least for the first 3 midpoints), thus justifying the use of an exponential decay term for the wave functions in the analytic theory. Nevertheless, the decay constant found for the tunneling current in this case, Fig. 5, differs for even or odd numbers of h-BN layers in the barrier. As already discussed in Section III, this difference arises from the overlap of the *lateral part* of the wave functions in the respective electrodes. This type of variation in the overlap of the lateral part of the wave functions is completely absent in the analytic theory; that theory fully includes the dependence on lateral wavevector difference between the electrodes (i.e. rotational alignment of the two graphene electrodes), but it does not describe any details of the barrier material nor of the alignment of the barrier with the graphene. Such "details" can substantially affect the prefactor for the tunneling and, to a lesser extent, the $\kappa$ value as well. In principle, the prefactor effects could be viewed as being wrapped up in the parameters $u_{11}$, $u_{12}$, and $D$ of the analytic theory. In the ideal case of fully aligned graphene electrodes and h-BN barrier layers, we find that the assumed values of unity for $u_{11}$ and $u_{12}$ in Ref. [6] were overestimates, by about a factor of 2 (maintaining a fixed $D$ value of 0.335 nm). This overestimate leads to a substantial error in the magnitude of the tunnel current since these values enter in the fourth power. However, for all realistic cases, i.e. with rotational and translational misalignment of the graphene and h-BN (and/or translational misalignment of the graphene electrodes themselves), then the overestimate of the tunneling current by the analytic theory is substantially greater. The first-principles theory of the present work provides a reliable, parameter-free method of obtaining a prefactor for the magnitude of the tunneling current, for all cases. On the other hand, the $\kappa$ values obtained in the first-principles theory (as presently implemented) are not so reliable, as we will now discuss.

Considering the 15× discrepancy found between our first-principles theory and experiment, we note that DFT consistently obtains band gaps that are too small, and hence will obtain $\kappa$ values that are too small. As an initial estimate of this effect, we consider the band gap value for h-BN from our GGA DFT computations. Referring to the band structures shown in Fig. S1, a gap of about 4.6 eV can be seen between the multiple bands that exist at the K point at energies near −1.4 and 3.2 eV (for the $\sqrt{7} \times \sqrt{7}$ and $\sqrt{19} \times \sqrt{19}$ cases some additional bands are seen split off from the CB, but a profusion of bands extending down from −1.4 eV and up from 3.2 eV is still apparent). This 4.6 eV applies to the 4-layer h-BN slabs that we employ in our computations; for computations of a full, periodically repeated h-BN bulk (that is, using the same lateral unit cell and arrangement of atoms as for the h-BN in our slab computations, but applying those to a bulk h-BN computation) we obtain band gaps that are about 0.15 eV smaller than those for the 4-layer slabs. The resultant bulk gap, 4.45 eV, is 1.55 eV less than the bulk h-BN value of 6.0 eV obtained both from experiment and from GW DFT.[35,36,37] Considering a two-band model for evanescent states, the maximal value of $\kappa$ along a complex-**k** loop connecting the conduction and valence bands varies like the square root of the band gap.[26] Hence, our 1.55 eV correction to the computed band gaps is equivalent to a $\sqrt{6 \text{ eV}/4.45 \text{ eV}} - 1 = 16\%$ correction to the $\kappa$ value, or an increase of 1.0 nm$^{-1}$. For 4 layers of h-BN, this correction would reduce the tunnel current by a factor of $\exp[\,2(1.0 \text{ nm}^{-1})(1.336 \text{ nm})] = 14$.

This estimated effect of the band gap correction yields a significant reduction in the tunnel current. To quantify this effect, a GW computation of the wave functions and resulting tunnel currents would be best; such a computation is, however, quite challenging. As an approximate alternative, we have utilized *strained* h-BN, with the strain providing an increase in the band gap. We achieve a strained K-point gap value near 6.0 eV for a situation with a 3×3 h-BN unit cell matched to a $\sqrt{7} \times \sqrt{7}$ graphene unit (19.11° rotational misalignment), corresponding to −13.2% strain in the h-BN. The geometrical arrangement and band structure for this situation are shown in Fig. S3, with a gap of 5.8 eV found in the band structure (for the 4-layer h-BN slab). Figure 9 shows the tunnel current for this situation, displaying a decay constant $\kappa$ of 6.7 nm$^{-1}$ (nearly no change in the prefactor $\eta$ is found compared to Fig. 8; only the $\kappa$ value is affected by the increase in the band gap). The tunnel current for 4 layers of h-BN is correspondingly reduced, and as seen in Fig. 9 the discrepancy between theory and experiment is now found to be a factor of only 2.5×. We therefore believe that we have achieved nearly quantitative agreement between theory and experiment for the magnitude of the G/h-BN/G tunneling current.

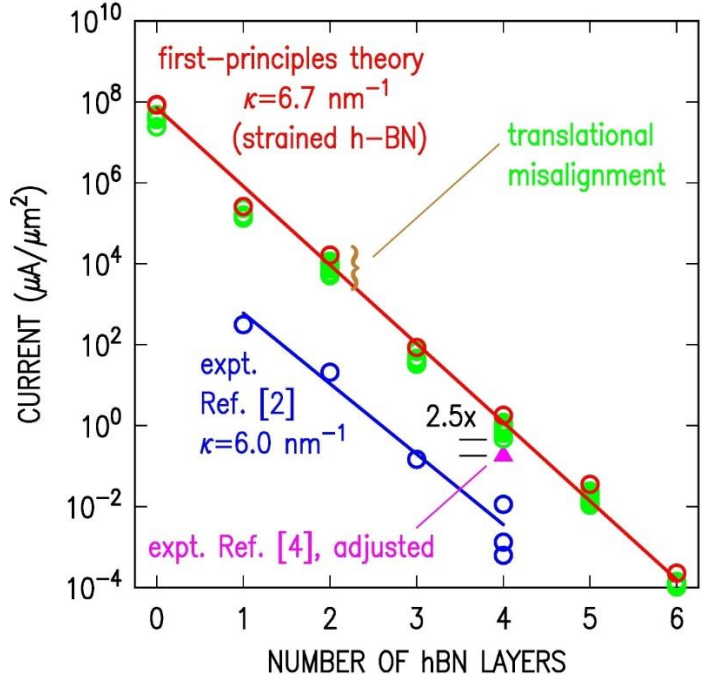

**FIG 9.** Theoretical results for strained h-BN on graphene (with theoretical band gap close to experimental value), compared to experimental results from references indicated. Rotational misalignment of 19.11° between the graphene and h-BN is used. For the theory, red data points indicate results for translationally aligned graphene electrodes, whereas green circles include translational misalignment of the electrodes. Lines show fits to $\eta \exp(-2\kappa d)$, as in Fig. 7. Comparing the theory (including misalignment) to the experiment of Ref. [4], a discrepancy of 2.5× occurs between them.

For our deduced decay constant of 6.7 nm$^{-1}$ in Fig. 9, this value is substantially greater than the 6.0 nm$^{-1}$ deduced the experimental data,[2] as shown in Figs. 8 and 9. However, two aspects of the data should be pointed out. First, some variation occurs in the values (especially for 4 h-BN layers, as seen in Figs. 8 and 9); from this, we estimate an uncertainty in the decay constant of *at least* ±0.3 nm$^{-1}$. Second, the data of Ref. [2] was obtained for a *non-resonant* condition, i.e. with uncontrolled rotation between the graphene electrodes (and between the graphene and the h-BN). For this situation, phonon participation in the tunneling process is known to occur.[38] If the probability of phonon-assisted tunneling increases with the thickness of the h-BN, then this would lead to a decrease in $\kappa$. For example, an increased phonon participation of a factor of 2 per h-BN layer would lead to a decrease in $\kappa$ of $\ln(2)/[2(0.334\text{ nm})] = 1.0$ nm$^{-1}$ for the non-resonant situation, which is quite substantial. Hence, we feel that our estimated $\kappa$ value of 6.7 nm$^{-1}$ for resonant tunneling is not inconsistent with the existing (non-resonant) experimental data.

Inclusion of the effects of rotational misalignment between graphene and h-BN, as described in Section III, played a modest role (~2× for 4 h-BN layers, although much more for 1 h-BN layer) in reducing the tunnel currents to the level found in experiment. Both Ge et al.[24] and Valsaraj et al.[25] have recently also examined rotational misalignment effects, and a comparison of our results with theirs is presented in Tables I and II. In the former, we list the ratios of the tunnel current for a 1×1 aligned situation (with translationally aligned graphene electrodes) compared to that for rotationally misaligned cases, for various thickness of the h-BN barrier. In the latter, we list ratios of the tunnel current for 1 h-BN layer compared to that for 2, 3, or 4 layers, for rotationally aligned and misaligned situations. Thus, all of the values in these two Tables represent factors by which the current *decreases*, due either to rotational misalignment (Table I) or to variation in the barrier thickness (Table II).

**TABLE I.** Comparison of theoretical results for tunnel current, as a ratio of the current for a 1×1 rotationally aligned (0°) case to that for rotationally misaligned situations.

| no. h-BN layers | this work | | | Valsaraj et al., Ref. [25] | | | Ge et al., Ref. [24] | | |
|---|---|---|---|---|---|---|---|---|---|
| | 1×1 0° | $\sqrt{7}\times\sqrt{7}$ 21.79° | $\sqrt{19}\times\sqrt{19}$ 13.17° | 1×1 0° | $\sqrt{7}\times\sqrt{7}$ 21.79° | $\sqrt{19}\times\sqrt{19}$ 13.17° | 1×1 0° | $\sqrt{7}\times\sqrt{7}$ 21.79° | $\sqrt{19}\times\sqrt{19}$ 13.17° |
| 1 | 1 | 9.3 | 13.7 | 1 | 8.7 | 25 | 1 | 147 | 71 |
| 2 | 1 | 4.3 | 3.1 | 1 | 8.7 | | 1 | | |
| 3 | 1 | 7.6 | 6.4 | 1 | 2.8 | | 1 | 250 | 58 |
| 4 | 1 | 1.7 | 1.7 | 1 | | | 1 | | |

**TABLE II.** Comparison of theoretical results for tunnel current, as a ratio of the current for a 1 h-BN layer compared to that for a specified no. of h-BN layers.

| no. h-BN layers | this work | | | Valsaraj et al., Ref. [25] | | | Ge et al., Ref. [24] | | |
|---|---|---|---|---|---|---|---|---|---|
| | 1×1 0° | $\sqrt{7}\times\sqrt{7}$ 21.79° | $\sqrt{19}\times\sqrt{19}$ 13.17° | 1×1 0° | $\sqrt{7}\times\sqrt{7}$ 21.79° | $\sqrt{19}\times\sqrt{19}$ 13.17° | 1×1 0° | $\sqrt{7}\times\sqrt{7}$ 21.79° | $\sqrt{19}\times\sqrt{19}$ 13.17° |
| 1 | 1 | 1 | 1 | 1 | 1 | 1 | 1 | 1 | 1 |
| 2 | 15 | 7.2 | 3.4 | 11 | 12 | | | | |
| 3 | 633 | 519 | 297 | 952 | 300 | | 105 | 178 | 87 |
| 4 | 62,000 | 11,300 | 7,500 | | | | | | |

Let us first examine the comparison in Tables I and II between our results and those of Valsaraj et al.[25] Both their work and our work employ first-principles wave functions obtained from DFT. However, Valsaraj et al. employ a quite different method than we do; rather than directly computing tunnel currents, they consider the splitting between the bonding and antibonding (symmetric and antisymmetric) states that form in the G/h-BN/G system. They take the square of this band splitting to be proportional to tunnel current, and they consider various rotational misalignment angles between the graphene and the h-BN. Examining the results of Table I, we find reasonable agreement (within a factor of 2) between results of our method and theirs. Similarly, in Table II we find agreement (to better than a factor of 2) between our results and those of Valsaraj et al. This overall agreement provides us with some confidence in the appropriateness of the theoretical methodology of the present work.

Turning now to the results of Ge et al. listed in Tables I and II, those workers have employed the NEGF method, using a tight-binding basis set for the electronic states of the graphene and h-BN. Examining the results in Table I, we find that Ge et al. obtain a much larger dependence of the tunnel current on rotational misalignment than either our work or that of Valsaraj et al. For the 21.79° misalignment the discrepancy is more than a factor of 10 between the results of Ge et al. compared to those of both us and Valsaraj et al., and for 13.17° the discrepancy is slightly less than a factor of 10. We believe that this discrepancy results from the rather approximate tight-binding description used by Ge et al. For the h-BN barrier material, 2 parameters (in addition to on-site energies) are used to model the bands, yielding a band structure in approximate agreement with that from first-principles theory.[25,39] Perhaps more significantly, a phenomenological term is used by Ge et al. to describe the variation in nearest neighbor interactions with lateral separation of the atoms.[25,40] Although this term has been used with some success to model twisted layers of graphene,[40] its applicability to twisted h-BN on graphene is less clear. Indeed, the large difference between the results of Ge et al. compared to the first-principles result of both our work and of Valsaraj et al. indicates, in our opinion, the rather qualitative nature of this tight-binding model for the case of h-BN on graphene.

**V. Summary**

In summary, we have performed a detailed comparison between theory and experiment for G/h-BN/G interlayer tunneling devices. A main goal of the study is to validate the first-principles DFT-Bardeen theory utilized for modeling the devices, so that similar computations involving other 2D materials such as TMDs or phosphorene can be accepted with confidence.[17] Compared to the experimental results of Mishchenko et al. for a G/h-BN/G device,[4] we find that tunnel currents computed with the Bardeen method employing first-principles GGA DFT wave functions are about 15× larger than experiment, so long as translational/rotational misalignment effects of the graphene and the h-BN (which produce a factor of ~5 reduction in the current) are included. Furthermore, it is demonstrated that the underestimation of the h-BN band gap in the DFT can play a significant role in further reducing the theoretical currents. Utilizing strained h-BN, with a band gap close to the experimental value, we obtain currents that agree with experiment to within a factor of 2.5×. Future computations, employing the GW method, are desirable for better quantification of the role of the band gap size.

One important issue that should be noted in connection with the theory for interlayer tunneling devices is that, in contrast to the experimental G/h-BN/G devices which had 4 layers of h-BN,[3] many of the recent theoretical results involve devices with perhaps only 1 or even 0 layers of h-BN.[11,12,13,17,21,22,23] The thin barriers yield large tunnel currents, as desirable for useful electronic devices. However, associated with the large currents, the effects of contact resistance can become quite significant. These might take the form of extrinsic effects, i.e. a simple voltage drop across the 2D leads and/or metallic junctions that connect to a device,[17,22] or in a more intrinsic way, as a continuously varying voltage drop along the source and drain materials themselves (which can significantly perturb the tunneling).[22,23,41] The former effect has been shown to be quite significant in recent experimental results,[42] and the latter effect can in principle be well treated with full computational packages that include three-dimensional electrostatic modeling as well as scattering in the electrodes.[21,22,23] Separately, another issue that can occur in interlayer tunneling devices is the formation of moiré (interference) patterns that form between lattice-mismatched layers of 2D material.[43] These patterns can lead to shifts in the band-edge energies across the area of the tunnel junction, hence restricting the achievement of steep subthreshold slope in a device. They also can lead to noticeable changes in the interlayer separation,[44] which in turn affects the tunnel current, as a function of position throughout the moiré unit cell. In the context of the present work, it is not clear how the *average* of the tunnel current across a unit cell would be affected by this variation in interlayer separation. These are among the issues that remain to be explored in future investigations of interlayer TFETs.

**Acknowledgements**

This work was supported by the Center for Low Energy Systems Technology (LEAST), one of six centers of STARnet, a Semiconductor Research Corporation program sponsored by Microelectronics Advanced Research Corporation (MARCO) and Defense Advanced Research Projects Agency (DARPA). Discussions with Di Xiao and Xiaoguang Zhang are gratefully acknowledged.

21

**Supplementary Information:**

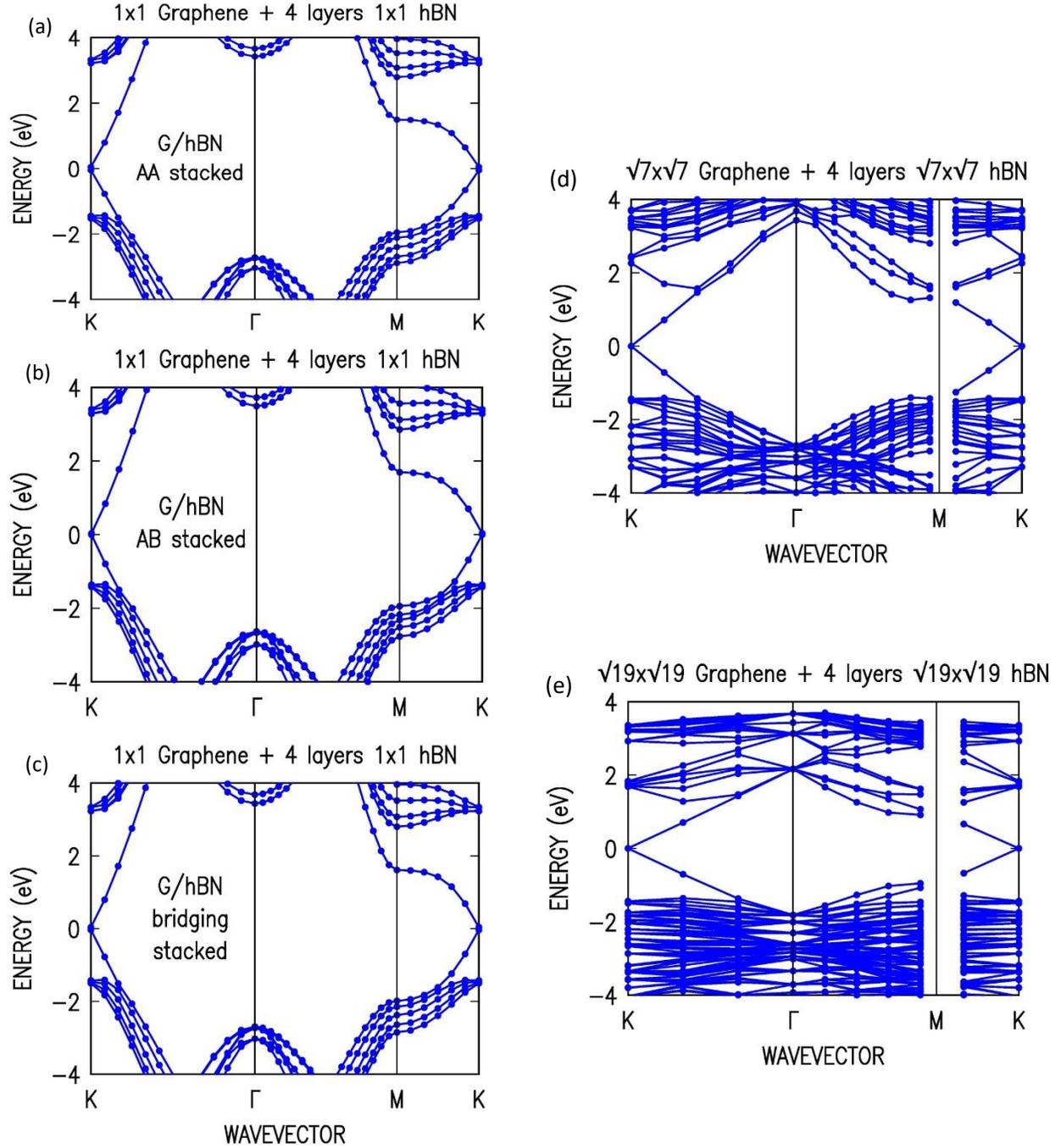

**FIG S1.** Band structures of graphene plus 4 layers of h-BN, with registration between the graphene and h-BN of (a) – (c) 1×1 (0° rotation) and with translationally aligned (AA), Bernal stacked (AB), and bridging stacked, respectively; (d) $\sqrt{7} \times \sqrt{7}$ (21.79° rotation) translationally aligned; and (e) $\sqrt{19} \times \sqrt{19}$ (13.17° rotation) translationally aligned. The size of the Brillouin zone varies in accordance with the unit cell size, with the K point having wavevector magnitude of (a) – (c) 17.0, (d) 6.41, and (e) 3.89 nm$^{-1}$. See Fig. 3 for detailed structural arrangements.

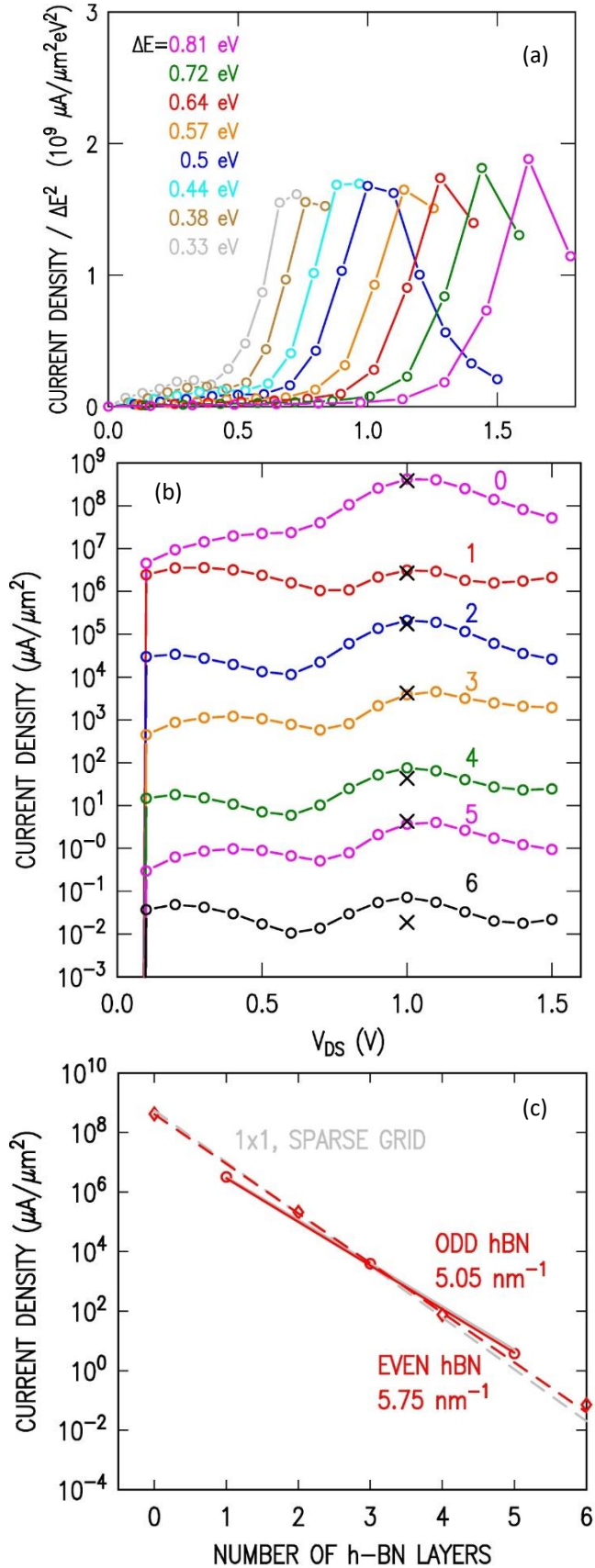

**FIG S2.** Results for 1×1 aligned arrangement of h-BN and graphene, using a 16× denser **k**-space grid than in the main text. (a) Current vs. source-drain bias voltage for 0 h-BN layers and for various assumed doping concentrations, specified by $\Delta E$, in the source and drain (to be compared with Fig. 5 of main text). (b) Current vs. source-drain voltage for various numbers of h-BN layers, as indicated, and doping of $\Delta E = 0.5$ eV. Values for the sparse **k**-space grid used in the main text are shown by the x-marks, multiplied by the correction factor of 0.71. (c) Current vs. number of h-BN layers, at source-drain voltage of 1.0 V and for doping of $\Delta E = 0.5$ eV. Values for the sparse **k**-space grid used in the main text (Fig. 6) are shown in gray, multiplied by the correction factor of 0.71, and results for the 16× denser grid are in red. Reasonably good agreement is found between results for the two grids, although for even numbers of h-BN layers the deduced value of $\kappa$ changes from 6.0 nm$^{-1}$ for the sparse grid (Fig. 6) to 5.75 nm$^{-1}$ for the fine grid. This same systematic trend of larger currents for the fine grid, for h-BN thicknesses of 4 and especially 6, is also apparent in (b).

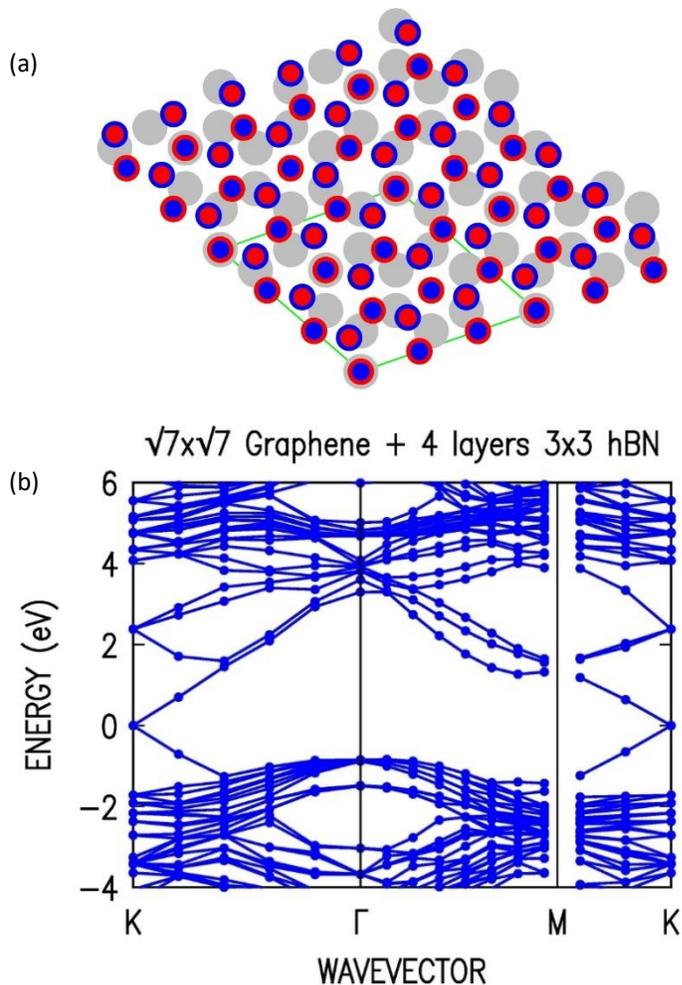

**FIG S3.** Results for structure with 3×3 unit cell of h-BN fit to $\sqrt{7} \times \sqrt{7}$ unit cell of graphene, and with a single C and B and/or N atom in each cell being translationally aligned. (a) Atomic arrangement, showing one ML of graphene (gray filled circles) and two MLs of h-BN (red and blue filled circles for B and N, respectively). Smaller circles correspond to atoms that are further above the graphene plane. Unit cell is shown in green. (b) Band structure (note the expanded energy scale compared to those in Fig. S1).